\documentclass[twocolumn,showpacs,aps,prl,superscriptaddress,letterpaper,nofootinbib]{revtex4}

\usepackage{graphicx}
\usepackage{dcolumn}
\usepackage{amsmath}
\usepackage{epsfig}

\input babarsym

\def\dpsignalpi {D^+ \rightarrow \pi^+\pi^0}
\def\dpsignalk {D^+ \rightarrow K^+\pi^0}
\def\dskks {D_s^+ \rightarrow K^+K^0_S}
\def\dpreference {D^+ \rightarrow K^-\pi^+\pi^+}

\newcommand{\BABARPubYear}    {06}
\newcommand{\BABARPubNumber}  {30}

\newcommand{\SLACPubNumber} {11861}
\newcommand{\LANLNumber} {0605044}
\newcommand{\lumi}    {124.3\invfb}
\newcommand{\yieldref} {101380}
\newcommand{\yieldrefstat} {415}
\newcommand{\yieldrefsys} {1374}
\newcommand{\effkpipi} {8.5}

\newcommand{\yieldpipi} {1229}
\newcommand{\yieldpipistat} {98}
\newcommand{\yieldpipisys} {10}
\newcommand{\effpipi} {7.8}
\newcommand{\fracpipi} {1.33}
\newcommand{\fracpipistat} {0.11}
\newcommand{\fracpipisys} {0.09}
\newcommand{\fracpipiexp} {\times 10^{-2}}
\newcommand{\bfpipi} {1.25}
\newcommand{\bfpipistat} {0.10}
\newcommand{\bfpipisys} {0.09}
\newcommand{\bfpipipdg} {0.04}
\newcommand{\bfpipiexp} {\times 10^{-3}}

\newcommand{\yieldkpi} {189.0}
\newcommand{\yieldkpistat} {35.2}
\newcommand{\yieldkpisys} {12.8}
\newcommand{\effkpi} {5.9}
\newcommand{\frackpi} {2.68}
\newcommand{\frackpistat} {0.50}
\newcommand{\frackpisys} {0.26}
\newcommand{\frackpiexp} {\times 10^{-3}}
\newcommand{\bfkpi} {2.52}
\newcommand{\bfkpistat} {0.47}
\newcommand{\bfkpisys} {0.25}
\newcommand{\bfkpipdg} {0.08}
\newcommand{\bfkpiexp} {\times 10^{-4}}

\def\figurebox#1#2#3{%
    \def\arg{#3}%
    \ifx\arg\empty
    {\hfill\vbox{\hsize#2\hrule\hbox to #2{\vrule\hfill\vbox to #1{\hsize#2\vfill}\vrule}\hrule}\hfill}%
    \else
    {\hfill\epsfbox{#3}\hfill}%
    \fi}

\long\def\inst#1{\par\nobreak\kern 4pt\nobreak
    {\it #1}\par\vskip 10pt plus 3pt minus 3pt}

\begin{document}

\bibliographystyle{prsty} 

\begin{flushleft}
hep-ex/\LANLNumber\\
SLAC-PUB-\SLACPubNumber\\
\babar-PUB-\BABARPubYear/\BABARPubNumber\\
\end{flushleft}

\title{
{\large \bf
Measurement of the $D^+ \rightarrow \pi^+\pi^0$ and $D^+ \rightarrow K^+\pi^0$ Branching Fractions } 
}

%
\author{B.~Aubert}
\author{R.~Barate}
\author{M.~Bona}
\author{D.~Boutigny}
\author{F.~Couderc}
\author{Y.~Karyotakis}
\author{J.~P.~Lees}
\author{V.~Poireau}
\author{V.~Tisserand}
\author{A.~Zghiche}
\affiliation{Laboratoire de Physique des Particules, F-74941 Annecy-le-Vieux, France }
% ...
\collaboration{The \babar\ Collaboration}
\noaffiliation

\date{\today}

\begin{abstract}
\noindent We present measurements of the branching fractions for the Cabbibo 
suppressed decays $D^+ \rightarrow \pi^+\pi^0$ and 
$D^+ \rightarrow K^+\pi^0$ based on a data sample corresponding to an integrated luminosity 
of $124.3$ fb$^{-1}$. The data
were taken with the \babar\ detector at the PEP-II $B$ Factory operating on and near the $\Upsilon(4S)$ resonance. 
We find  
${\cal B}(D^+ \rightarrow \pi^+\pi^0) = ( 1.25 \pm 0.10 \pm 0.09 \pm 0.04 ) \times 10^{-3}$ 
and 
${\cal B}(D^+ \rightarrow K^+\pi^0) = ( 2.52 \pm 0.47 \pm 0.25 \pm 0.08 ) \times 10^{-4}$, 
where the first uncertainty is statistical, the second systematic and 
the last error is due to the uncertainties in the absolute branching fraction scale for $D^+$ mesons.
This represents the first observation of the doubly Cabibbo-suppressed 
$D^+ \rightarrow K^+\pi^0$ decay mode and 
a new measurement of the $D^+ \rightarrow \pi^+\pi^0$ branching fraction.
\end{abstract}


\maketitle
Measurements of rare hadronic $D^+$ decays provide insight into $SU(3)$ flavor symmetry, 
QCD dynamics, and weak flavor mixing \cite{Chau:1993ec}. 
Studies of these decays are useful for $D^0 \bar{D^0}$ mixing analyses, 
which benefit from improved measurements of $D^+ \rightarrow \pi^+\pi^0$ and 
$D^+ \rightarrow K^+K^0$ branching fractions in order
to understand the size of the $SU(3)$-violating  effects in $D$ meson decays. 
In addition, doubly Cabibbo-suppressed decays such as $D^0 \rightarrow K^+\pi^-$ 
complicate measurements of flavor oscillations in hadronic $D^0$ decays. 
Knowledge of the $SU(3)$-related channels reported here can lead
to a better understanding of this background.
Previous analyses of these $D^+$ decays were reported by MARK III, CLEO, 
and FOCUS \cite{Baltrusaitis:1985we,Arms:2003ra,Link:2004vk}.

This analysis is based on data recorded with the \babar\ detector at the \pep2\ asymmetric-energy \epem storage ring at the 
Stanford Linear Accelerator Center. The data sample corresponds to an integrated luminosity of \lumi recorded at center-of-mass (CM) energies
$\sqrt{s} = 10.58 \gev$ and $10.54 \gev$ and includes approximately $167 \times 10^6$ $e^+e^- \rightarrow c\bar{c}$ events.

The \babar\ detector is described in detail elsewhere~\cite{detector}. Charged particle momenta are measured with a 5-layer double-sided silicon 
vertex tracker (SVT) and a 40-layer drift chamber (DCH), both inside the 1.5 T magnetic field of a superconducting solenoid. A calorimeter (EMC) consisting of 
6580 CsI(Tl) crystals measures electromagnetic energy. Charged hadron identification is provided by measurements of the rate of ionization 
energy loss, $dE/dx$, in the tracking system and of the Cherenkov angle obtained from a ring-imaging Cherenkov detector (DIRC). 
The instrumented flux return of the magnet allows discrimination of muons from pions. 

We use a Monte Carlo simulation of the \babar\ detector based on \geant4~\cite{geant4} 
to validate the analysis and to determine the reconstruction efficiencies for the two signal modes $\dpsignalpi$ and $\dpsignalk$,
as well as for the decay $\dpreference$ \footnote{Unless explicitly stated, charge conjugate reactions are implicitly included throughout this paper.}, 
which is used as a reference to normalize our results.
Simulated events are generated with the Pythia event generator \cite{Sjostrand:2000wi}. 

We reconstruct $D^+$ meson candidates in the signal modes by combining a charged track, identified either as a pion or kaon, with a reconstructed 
$\pi^0$ candidate.
Until better knowledge of the overall $e^+e^- \rightarrow D^{+}X$ production rate is obtained, any measurement of
absolute $D^+$ branching fractions will be limited by the uncertainty in the number of $D^{+}$ mesons in the data sample. 
We avoid this uncertainty by measuring our signal modes relative to the high statistics, well-measured $\dpreference$ decay mode.

In order to reduce the large amount of combinatorial background in the $D^{+}$ signal modes, 
we include
only $D^{+}$ mesons that originate from $D^{*+} \rightarrow D^{+} \pi^{0}$ decays.
To minimize systematic uncertainties in the reconstruction
of the low momentum $\pi^{0}$ from the $D^{*+}$ decay,
this is done for both the signal and reference channels 

Only events with at least three charged tracks are selected for this analysis. 
Charged tracks are required to have a distance of closest approach to the 
interaction point in the plane transverse to the beam axis of less than 1.5~cm, a distance of closest approach 
along the beam direction of less than 10~cm, a minimum transverse momentum of $100\,\mevc$, and at least 12 DCH hits. 
All candidate tracks in the reconstructed decay chains must satisfy a set of pion or kaon identification criteria based
on the response of the DIRC and the $dE/dx$ measurements in the tracking system.

A pair of energy clusters in the EMC, which are isolated from any charged tracks and have the expected lateral 
shower shape for photons, is considered a $\pi^0$ candidate if both clusters exceed 30 MeV, and the 
associated invariant mass of the pair is between 0.115 \gevcc and 0.150 \gevcc. The energy of the 
$\pi^{0}$ candidate in the laboratory frame is required to be greater than $0.2 \gev$. 

We accept a $D^+$ candidate if its invariant mass falls between $1.7 \gevcc$ and $2.0 \gevcc$. 
In addition we require that the cosine of the
helicity angle, $\theta_{h}$ 
\footnote{We define $\theta_h$ as the angle between the direction of the charged daughter particle of the $D^+$ decay and the direction of the $D^{*+}$
meson evaluated in the $D^+$ rest frame.},
which is uniformly distributed for signal events but peaks at $\pm 1$ for background, 
satisfies $-0.9 < \cos(\theta_{h}) < 0.8$ for the $\dpsignalpi$ mode 
and $-0.9 < \cos(\theta_{h}) < 0.7$ for the $\dpsignalk$ mode.

Two positively charged pion tracks and a negatively charged 
kaon track are combined in a vertex fit to form a
$\dpreference$ candidate. We require the chi-squared probability of the vertex fit to be 
${\cal P} > 0.001$. The candidate is accepted 
if the invariant mass of the $D^+$ lies between 1.75 \gevcc and 1.95 \gevcc. The smaller range 
compared with the signal modes
 reflects the better resolution for this decay mode, which has only charged tracks in the final state.

$D^+$ candidates are combined with a reconstructed $\pi^0$ to select 
$D^{*+} \rightarrow D^{+} \pi^{0}$ decays. An additional requirement on the center of mass 
momentum, 
$p_{CM} < 0.45 \gevc$, is applied to $\pi^0$
candidates used in the $D^{*+}$ reconstruction.
Only $D^{*+}$ candidates with a mass difference $\Delta m = m_{D^{*+}} - m_{D^+}$ less than
$0.155 \gevcc$ 
are accepted for this analysis. A requirement on the normalized 
momentum of the $D^{*+}$ meson\footnote{
The normalized momentum $x_{D^*}$ is defined as
$x_{D^*}\, =\, p_{D^{*}}^{CM}/\sqrt{s/4 - m_{D^*}^2}$
where $p_{D^*}^{CM}$ is the momentum of the $D^*$ meson in the CM frame and $s$ is the square of the
energy of the initial $e^+e^-$ system.
},
$x_{D^*} > 0.6$, corresponding to a $D^*$ center of mass momentum greater than 2.9 $\gevc$, 
eliminates backgrounds from $B$ meson decays and further reduces the combinatorial background.
If more than one $D^{*+}$ candidate is reconstructed in an event, we choose the one with the larger $x_{D^*}$ value.

With these requirements applied to Monte Carlo events, we obtain reconstruction efficiencies 
of \effpipi\% for the $\dpsignalpi$ mode, \effkpi\% for the $\dpsignalk$ 
mode, and \effkpipi\% for the $\dpreference$ mode. 
 
We extract the signal yield for each of the three decay modes from the invariant mass distribution of the $D^+$ candidates with
unbinned extended maximum likelihood fits.

For the reference mode, the remaining background in the $K^-\pi^+\pi^+$ invariant mass distribution is described by a first order
polynomial; the signal lineshape is modeled by a double Gaussian function.
In order to accommodate possible differences in the $D^+$ momentum distribution in data and Monte Carlo events, 
a weight function describing the relative change in reconstruction efficiency as a 
function of the $D^+$ momentum 
is included in the likelihood function.
Similar corrections are applied to the signal mode fits described below. 
A second weight function is used in the $\dpreference$ fit to correct for potential 
differences between the $K^-\pi^+\pi^+$ Dalitz plot structure in data and simulated events.

\begin{figure}[tb]
\begin{center}
\begin{picture}(1,210)
\thicklines
\put(-100,0){\includegraphics[scale=0.4]{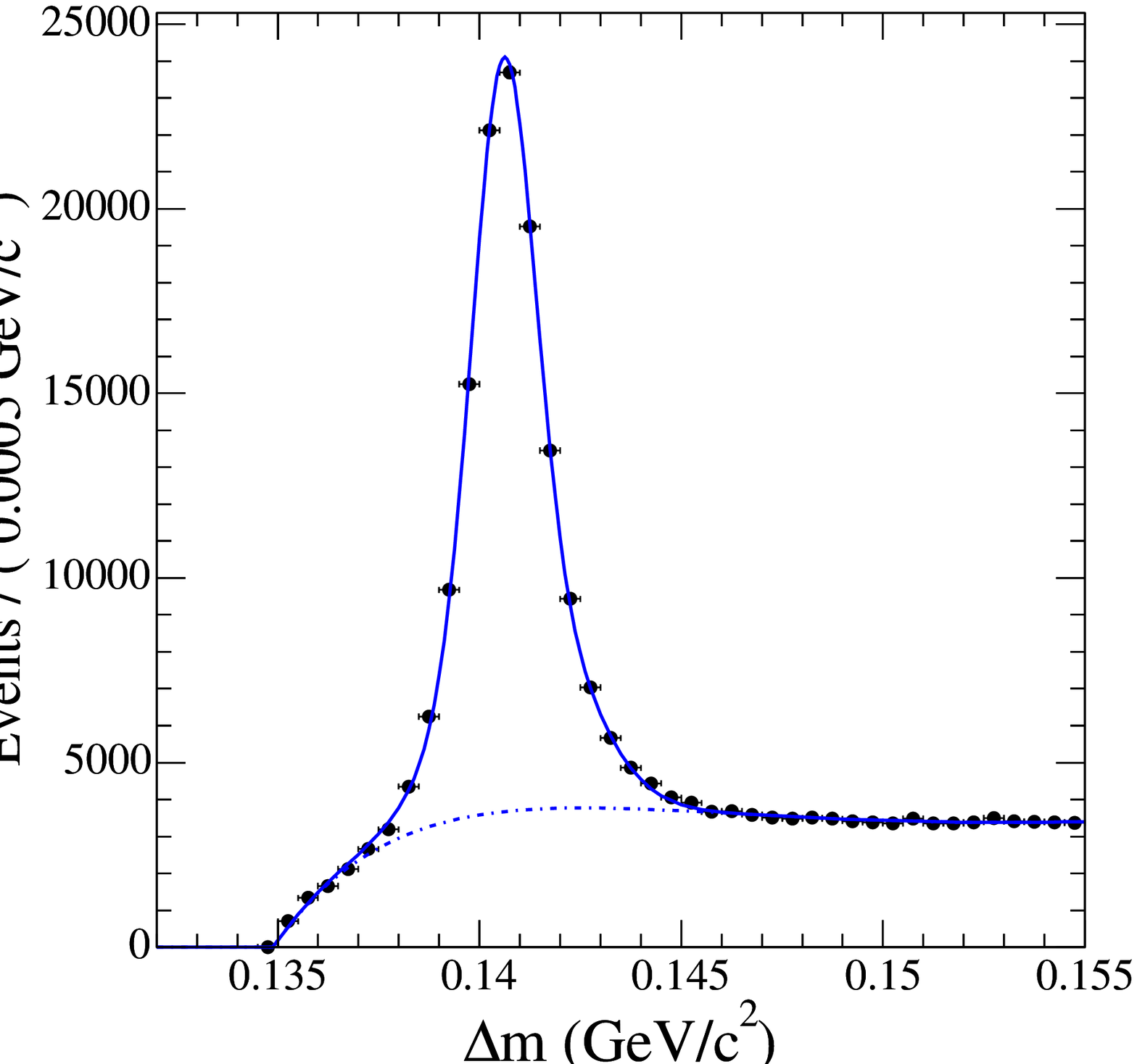}}     
\put(-20,21){\line(0,1){60}}
\put(12,21){\line(0,1){60}}
\put(42,21){\line(0,1){60}}
\put(80,120){\makebox(0,0)[tr]{Sideband}}
\put(-18,120){\makebox(0,0)[tr]{Signal}}
\put(-40,108){\vector(1,-2){28}}
\put(50,108){\vector(1,-2){28}}
\end{picture}
\caption[$\Delta m$ Signal and Sideband]{\label{fig:deltammc}$\Delta m$ distribution
for simulated $D^{*+} \to D^+\pi^0$, $\dpreference$ events.
The vertical lines mark the signal and sideband regions defined in the text.
The solid line shows the result of the fit; the background is given by the dash-dotted line.}
\end{center}
\end{figure}

Not all of the $D^+$ candidates in our sample originate from $D^{*+} \rightarrow D^+ \pi^0$ decays.
Some $D^+$ mesons from other 
sources can combine with a random $\pi^0$ in the event to pass the $\Delta m$ requirement.
This background is uniformly distributed in $\Delta m$;
in the invariant mass distribution it peaks at the $D^+$ mass.
While it might be expected that this effect cancels when we calculate branching fraction ratios between signal and
reference modes, Monte Carlo studies indicate that this is not the case. 
A correction, extracted from data, must be applied to compensate for the difference
in relative efficiencies between this peaking $D^+$ background in signal and reference
modes, caused mostly by the helicity angle requirement.
We use the $\Delta m$ sideband, shown in Fig.~\ref{fig:deltammc}, 
to determine the corrected $D^+$ yield. 
The $\Delta m$ signal region is defined as a $2\sigma$ window around the nominal $D^{*+} - D^+$ mass difference; the sideband extends from
$5\sigma$ above the nominal value to $0.155 \gevcc$.
The signal contains $D^+$ decays from all sources, while $D^+$ mesons present in the $\Delta m$ sideband come
from sources other than $D^{*+} \rightarrow D^+ \pi^0$ decays. 
The invariant mass distributions for the signal and sideband
regions are fitted simultaneously with identical signal shapes. The yield in the sideband is constrained 
to be greater than or equal to zero to avoid an unphysical enhancement of the signal yield should
the peaking $D^+$ background fluctuate low.
We scale the $D^+$ yield from the sideband by the ratio of the 
integrals under the combinatorial background curve in the signal and sideband regions and subtract this value from the yield in the
signal region to extract the net yield of $D^{*+} \rightarrow D^+ \pi^0$, $\dpreference \ (\pi^+\pi^0,\ K^+\pi^0)$ decays corrected for the peaking background.

For the $\dpsignalpi$ and $\dpsignalk$ signal modes,
the double Gaussian signal function is replaced by
a bifurcated Gaussian function, which gives a better description of the increased width of the signal toward lower
masses, caused by radiative losses associated with $\pi^0$ reconstruction.
An exponential function is added to the linear background parameterization to model backgrounds from misreconstructed decays 
such as $D^+ \rightarrow K^0_S K^+ \;\mbox{with}\;  K^0_S \rightarrow \pi^0 \pi^0$ or 
$D^0 \rightarrow K^0_S \pi^0 \;\mbox{with}\;  K^0_S \rightarrow \pi^+ \pi^-$, where two of the 
three decay products are used to reconstruct a signal mode candidate. 
This background contribution peaks at low mass, but has a long tail that extends into the $D^+$ signal region. 

The $\dpsignalk$ mode has an additional background.
Decays of $D_s^+$ mesons to $K^+K^0_S$ final states with the $K^0_S$ decaying to two neutral pions can cause an excess 
in the invariant $D^+$ mass distribution
if they are mistakenly reconstructed as $D^+ \rightarrow K^+ \pi^0$, and are combined with a random $\pi^0$ to mimic a $D^{*+}$ signal.
Based on a Monte Carlo study of $D^+_s \rightarrow K^+ K^0_S$ events we model this additional background component with a
Gaussian function centered at $1.807 \gevcc$ and with a width of $0.028 \gevcc$.

Signal and background shapes used in the fits are derived from Monte Carlo events. We minimize
systematic uncertainties due to differences between data and the simulation by allowing most parameters to vary in the data fits.
The only exceptions are in the $\dpsignalk$ mode where the expected yield is too small to determine the signal shape parameters
directly from data. Instead, we use the parameters found in the $\dpsignalpi$ data fit 
with the widths reduced by 5\%. This
correction was obtained in a Monte Carlo study of $\dpsignalk$ and $\dpsignalpi$ events. The second exception is the shape
of the $D^+_s \rightarrow K^+K^0_S$ background which is constrained to values obtained from Monte Carlo simulations.

\begin{figure}[tb]
\begin{center}
\epsfig{file=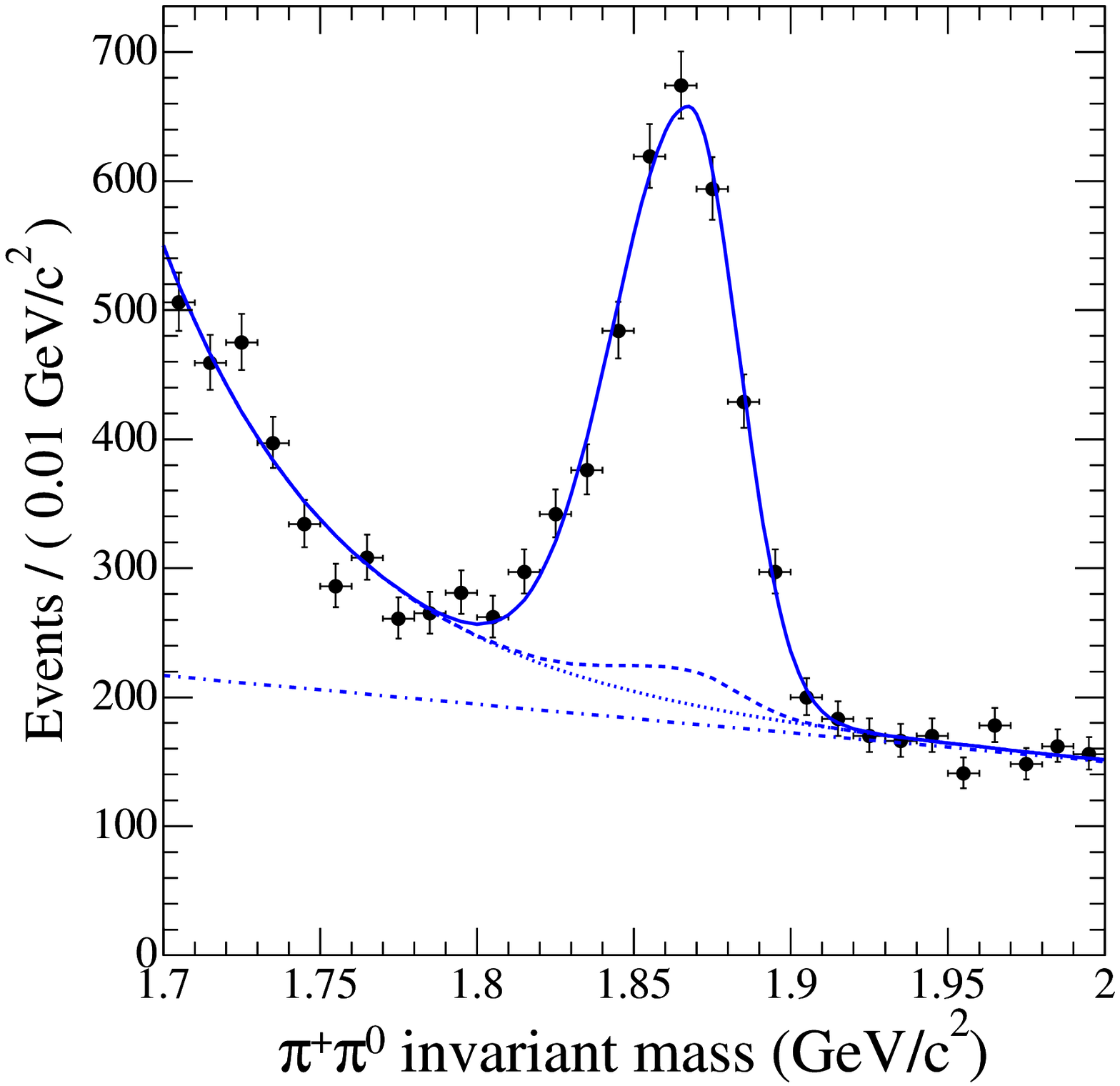,width=0.4\textwidth}
\caption[Full Monte Carlo Fit $\dpsignalpi$]{\label{fig:fullmcfitpipi} $\pi^+\pi^0$ invariant mass distribution 
in the $\dpsignalpi$ signal mode for simulated data. 
The background is modeled by a first order polynomial (dash-dotted) and an exponential (dotted), 
while the signal is fitted by a bifurcated Gaussian function. 
The dashed line shows the peaking $D^+$ background determined from the $\Delta m$ sideband fit.}
\end{center}
\end{figure}

A Monte Carlo sample corresponding to an integrated luminosity of approximately 80~fb$^{-1}$ is used to validate the fit procedure. The
branching fraction ratio of $\dpsignalpi$ to $\dpreference$ decays in the simulation 
is $2.8\times 10^{-2}$. With the yields, $N_{fit}$, extracted from the $\dpsignalpi$ and $\dpreference$ fits 
and the previously determined reconstruction efficiencies $\epsilon$, we obtain
\begin{eqnarray}
\frac{{\cal B}(\dpsignalpi)}{{\cal B}(\dpreference)} & = &
\frac{N_{fit}(\pi^+\pi^0) \cdot \epsilon_{K^-\pi^+\pi^+}}{N_{fit}(K^-\pi^+\pi^+) \cdot \epsilon_{\pi^+\pi^0}}\\
 & = & (2.7 \pm 0.1) \times 10^{-2}
\label{eq:ratio}
\end{eqnarray}
in good agreement with the expected value. 
As an example, the $\dpsignalpi$ fit for simulated data and the different fit components are shown in Fig.~\ref{fig:fullmcfitpipi}.
We repeat this study for the $\dpsignalk$ mode with the branching fraction ratio in the
Monte Carlo sample set to $2.8 \times 10^{-3}$ and find a value of $(3.1 \pm 0.6) \times 10^{-3}$ 
for the ratio of the efficiency-corrected yields returned by the $\dpsignalk$ and $\dpreference$ fits.

\begin{figure*}[tb]
\begin{center}
\begin{minipage}{0.3\textwidth}
 \begin{center}
  \begin{picture}(1,170)
    \put(-80,0){\includegraphics[width=1.1\textwidth]{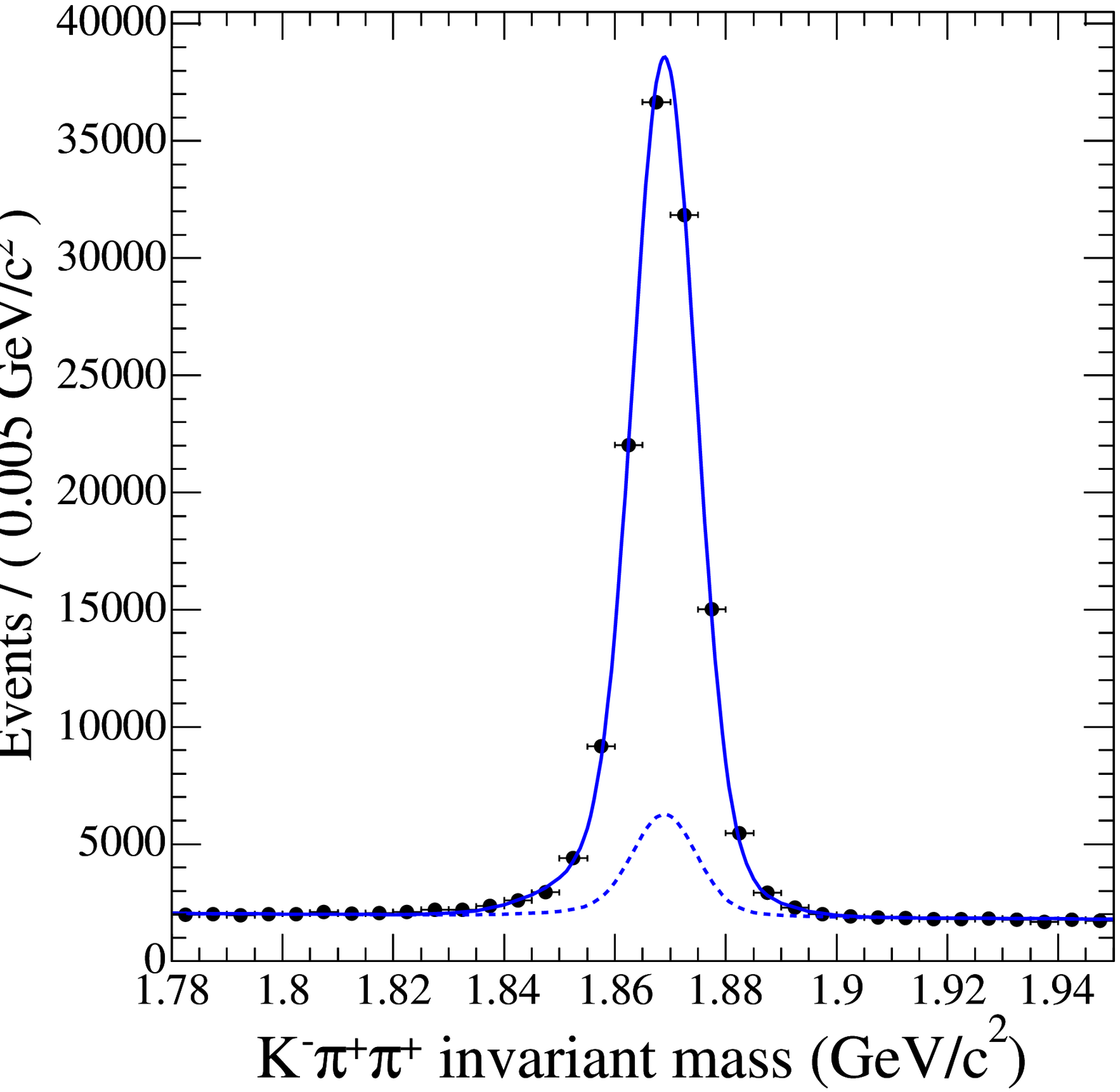}}     
    \put(60,135){\makebox(0,0)[tr]{(a)}}
  \end{picture}
 \end{center}
\end{minipage}
\begin{minipage}{0.3\textwidth}
 \begin{center}
  \begin{picture}(1,170)
    \put(-80,0){\includegraphics[width=1.1\textwidth]{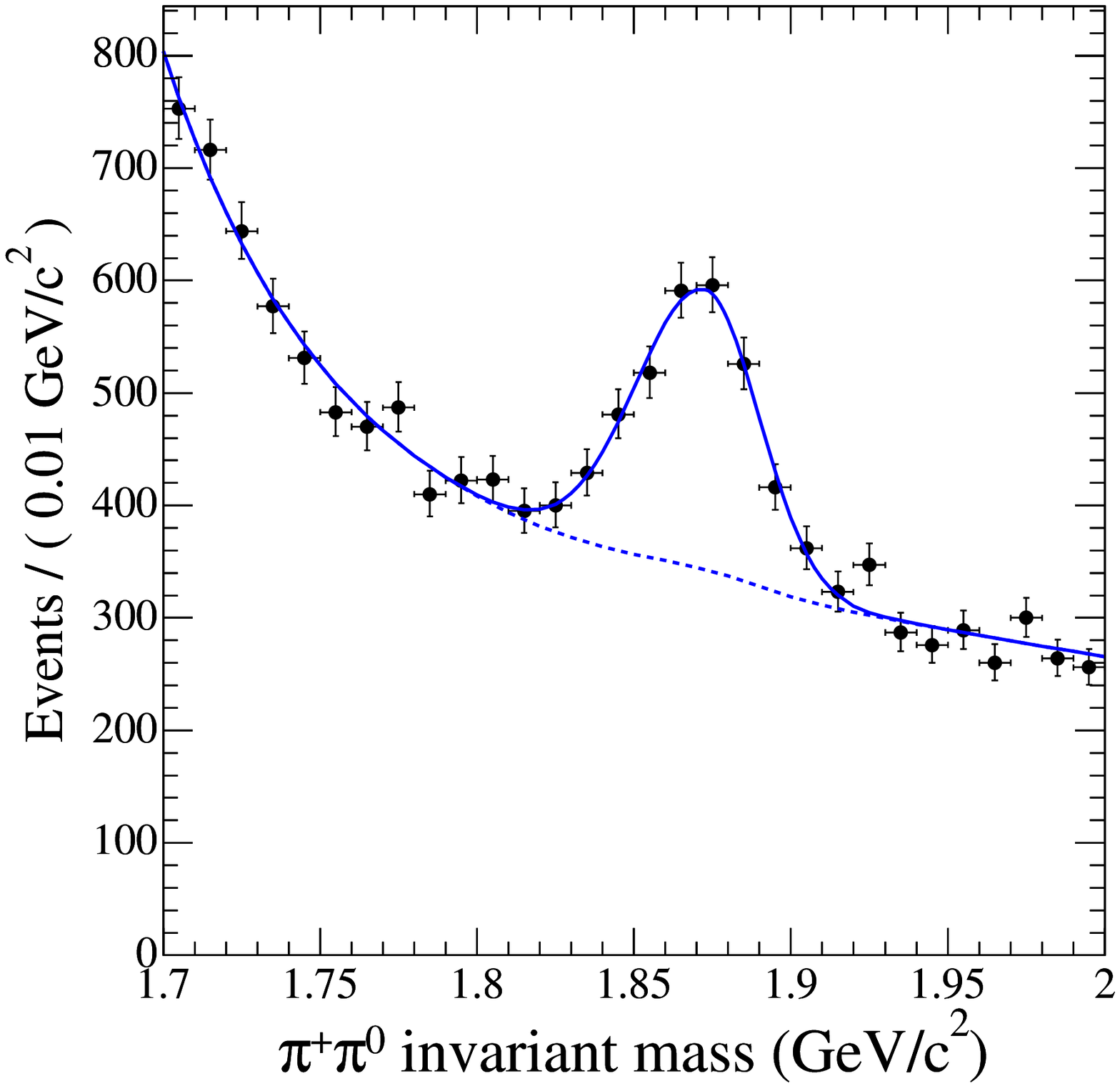}}     
    \put(60,135){\makebox(0,0)[tr]{(b)}}
  \end{picture}
 \end{center}
\end{minipage}
\begin{minipage}{0.3\textwidth}
 \begin{center}
  \begin{picture}(1,170)
    \put(-80,0){\includegraphics[width=1.1\textwidth]{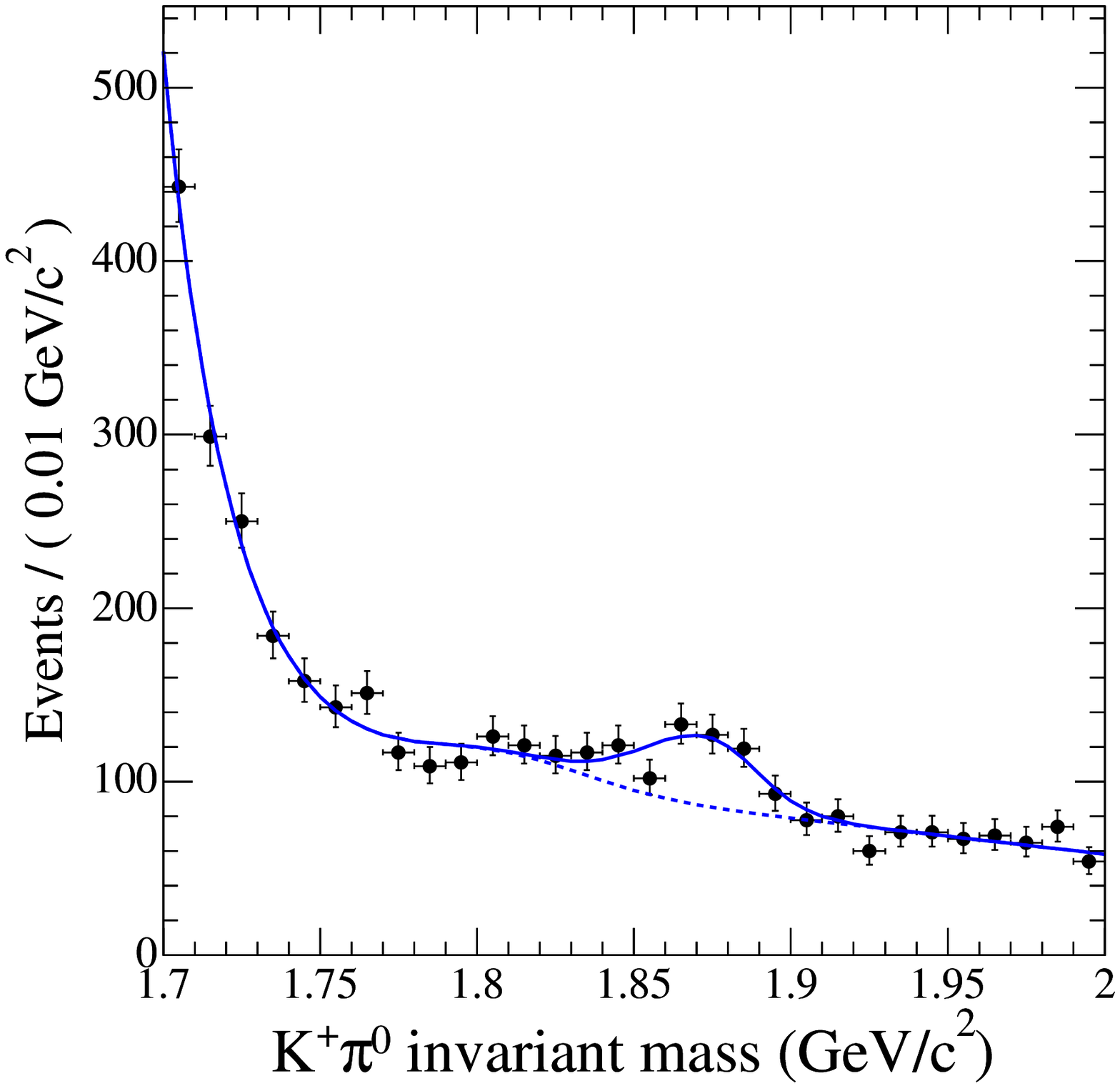}}     
    \put(60,135){\makebox(0,0)[tr]{(c)}}
  \end{picture}
 \end{center}
\end{minipage}
\caption[Run 1-3 Data Fits]{\label{fig:fulldata}Likelihood fit results for the full data sample for (a)
$\dpreference$, (b) $\dpsignalpi$, and (c) $\dpsignalk$ decays. The dashed lines show the projected backgrounds in the
signal region.
}
\end{center}
\end{figure*}

Figure \ref{fig:fulldata} shows the fit results for the full data sample. 
The signal yields are $N_{fit}(K^-\pi^+\pi^+) = \yieldref \pm \yieldrefstat$ for the
reference mode,
$N_{fit}(\pi^+\pi^0) = \yieldpipi \pm \yieldpipistat$  for the
$\dpsignalpi$ signal mode and
$N_{fit}(K^+\pi^0) = \yieldkpi \pm \yieldkpistat$ for the $\dpsignalk$ signal mode. The errors
are statistical only.

The systematic errors in this analysis include uncertainties in the reconstruction
efficiencies as well as the errors associated with the event yields returned by the maximum likelihood fit.
Since we measure the branching fraction for $\dpsignalpi$ and $\dpsignalk$ relative to
the $\dpreference$  reference mode several systematic uncertainties in the efficiencies cancel or are
reduced. Individual systematic contributions listed in the following paragraph apply to the measurement
of the ${\cal B}(\dpsignalpi)$ to ${\cal B}(\dpreference)$ ratio. 
Uncertainties of the ${\cal B}(\dpsignalk)$ to ${\cal B}(\dpreference)$ ratio 
are indicated in parentheses if they are different.

The relative systematic error on the efficiencies
includes contributions of 1.9\% from charged track reconstruction and vertexing,
0.9\% (0.4\%) from particle identification, 3.2\% due to 
uncertainties in $\pi^0$ reconstruction,
and 1.1\% (1.2\%) because of limited Monte Carlo statistics.
Uncertainties in reconstruction efficiency of the slow $\pi^0$ from $D^{*+}$ decays cancel in the
branching fraction ratios.
However, because of the $\pi^0$ in the final state of the signal modes the $D^+$ mass resolution differs
between signal and reference modes, which has an
effect on the shape of the peak in the $\Delta m$ distribution. Based on a study of simulated events
we assign a 5\% systematic error due to this effect.
The difference between the $\Delta m$ shapes in data and Monte Carlo events leads to an additional systematic
uncertainty of 1.4\% for the efficiency ratio. We vary the signal and background parameterizations used in
the fits of the $D^+$ invariant mass distributions 
and derive a systematic uncertainty of 1.5\% for the $\dpsignalpi$ to $\dpreference$ 
efficiency ratio. The fixed signal shape used to determine the amount of peaking $D^+$ background in the
$\dpsignalk$ signal mode causes an additional systematic uncertainty that we estimate by varying the
parameters of the Gaussian fit function within their errors and refitting the data. The quadrature sum of
both effects gives a 2.7\% systematic uncertainty in the $\dpsignalk$ to $\dpreference$ efficiency
ratio.

We add the individual contributions in quadrature and obtain a total systematic uncertainty for the
$\dpsignalpi$ to $\dpreference$ and $\dpsignalk$ to $\dpreference$ efficiency ratios of  
6.7\% and 7.0\%, respectively.

The various weights, as well as the signal to sideband scale factor used in the maximum likelihood
fits, give rise to systematic uncertainties in the $D^+$ yield. We study effects caused by the 
weight functions
by varying their parameterizations within their errors, and find that this
leads to a systematic error of 0.8\% for the two signal modes and 1.2\% for the $\dpreference$ reference mode.

The scale factor used for the peaking $D^+$ background correction, extracted from data using the
$\Delta m$ distribution in the reference mode, is 
found to be $(34.5 \pm 1.6)$\%. The same scale factor is used for all
three $D^+$ decay modes which results in an additional systematic error of 1.5\%. 
For the $\dpsignalpi$ and $\dpreference$ modes, we combine this in quadrature
with the statistical error of the scale factor to
obtain a total systematic uncertainty
in the amount of peaking $D^+$ background of 4.9\%.
In the $\dpsignalk$ sample, the $D^+$ yield in the sideband region fluctuates to a
negative value; thus the nominal data fit constrains this component to zero. To estimate
the uncertainty due to the peaking $D^+$ background in this mode we repeat the fit without
the constraint and find a difference in signal yield between the constrained and the unconstrained
fit of 9.5 events. 
We use this value as the systematic error 
on the background-corrected $\dpsignalk$ yield.

An additional systematic uncertainty in the $\dpsignalk$ analysis is due to the
$D_s^+ \rightarrow K^+K^0_S$ background component.
Based on the measured branching fraction \cite{Eidelman:2004wy} for this decay 
we expect $87 \pm 58$ $\dskks$ background events in our data sample. 
This is consistent with the fitted yield of $118 \pm 53$.
When we vary the parameterization of the $\dskks$ background function in the fit,
the reconstructed $\dpsignalk$ yield changes by 4.5\%, 
which is taken into account as a systematic error.

With these systematic errors combined in quadrature, the yield in the reference mode is
$N_{\dpreference} = \yieldref \pm \yieldrefstat \pm \yieldrefsys $. For the signal modes we find
$N_{\dpsignalpi} = \yieldpipi \pm \yieldpipistat \pm \yieldpipisys $ and 
$N_{\dpsignalk} = \yieldkpi \pm \yieldkpistat \pm \yieldkpisys $.

Following Eq. \ref{eq:ratio}, we combine these measurements with the reconstruction efficiencies to obtain
\begin{equation*}
\frac{{\cal B}(\dpsignalpi)}{{\cal B}(\dpreference)} = ( \fracpipi \pm \fracpipistat \pm \fracpipisys ) \fracpipiexp
\end{equation*}
and
\begin{equation*}
\frac{{\cal B}(\dpsignalk)}{{\cal B}(\dpreference)} = ( \frackpi \pm \frackpistat \pm \frackpisys ) \frackpiexp.
\end{equation*}
Using ${\cal B}(\dpreference) = 0.094 \pm 0.003$ which is the weighted average
of a recent CLEO-c result \cite{He:2005bs} and the PDG 
value \cite{Eidelman:2004wy}, 
we derive the branching fractions for the two signal modes
\begin{equation*}
{\cal B}(\dpsignalpi) = ( \bfpipi \pm \bfpipistat \pm \bfpipisys \pm \bfpipipdg ) \bfpipiexp
\end{equation*}
and 
\begin{equation*}
{\cal B}(\dpsignalk) = ( \bfkpi \pm \bfkpistat \pm \bfkpisys \pm \bfkpipdg ) \bfkpiexp,
\end{equation*}
where the last error is due to the experimental uncertainty in the $\dpreference$ branching fraction.
We compute the significance ${\cal S}$ of the $\dpsignalk$ signal as
${\cal S} = \sqrt{2(ln{\cal L}(N_s) - ln{\cal L}(N_s = 0))}$, where ${\cal L}(N_s)$ is the maximum likelihood at
the nominal fit yield, and ${\cal L}(N_s = 0)$ is the value of the likelihood for $N_s = 0$.
We include systematic uncertainties by repeating this procedure while varying the fit parameters within their errors.
The smallest signal significance obtained in this manner 
is 6.5 standard deviations.

This represents the first observation of the doubly Cabibbo-suppressed $\dpsignalk$ decay mode, and 
a new measurement of the $\dpsignalpi$ branching fraction.

We can compare our results to theoretical expectations and evaluate the size of $SU(3)$ violation in these
decays. 
In the limit of $SU(3)$, the ratio 
$$
R_{SU(3)} = 2 \left| \frac{V_{cs}}{V_{cd}}\right| ^2
\frac{\Gamma(\dpsignalpi)}{\Gamma(D^+ \rightarrow \bar{K}^0 \pi^+)}
$$
is expected to approach unity \cite{Chau:1993ec}.
The extra factor of 2 arises because of the normalization of the $\pi^0$ wavefunction.
Combining values for the CKM matrix elements and 
${\cal B}(D^+ \rightarrow \bar{K}^0 \pi^+)$ \cite{Bigi:1994aw}
taken from \cite{Eidelman:2004wy} with our result for $\dpsignalpi$, we find
$R_{SU(3)} = 1.54 \pm 0.27$. 
At the decay amplitude level this corresponds to a 25\% deviation from 
$SU(3)$ symmetry, consistent with theoretical expectations \cite{Savage:1991wu}.

We can also compare doubly Cabibbo-suppressed decays of charged and neutral $D$ mesons. The two decays $\dpsignalk$
and $D^0 \rightarrow K^+ \pi^-$ differ only in the flavor of the spectator quark in
the $D$ meson. In the absence of $D^0$ mixing and taking into account that the $D^+$ decay includes a $\pi^0$ in the final state,
the ratio of decay rates is expected to be
1/2. This ratio could be modified by W-annihilation and W-exchange amplitudes that 
contribute differently to $\dpsignalk $ and $D^0\rightarrow K^+\pi^-$ decays \cite{Chiang:2001av}. 
Experimentally we find
$$
\frac{\Gamma(\dpsignalk)}{\Gamma(D^0\rightarrow K^+\pi^-)} =
\frac{{\cal B}(\dpsignalk)}{{\cal B}(D^0\rightarrow K^+\pi^-)}\frac{\tau_{D^0}}{\tau_{D^+}} =
0.71 \pm 0.16
$$
where the values for the $D$ lifetimes and ${\cal B}(D^0 \rightarrow K^+\pi^-)$ are taken 
from \cite{Eidelman:2004wy}. 
We are grateful for the excellent luminosity and machine conditions
provided by our \pep2\ colleagues, 
and for the substantial dedicated effort from
the computing organizations that support \babar.
The collaborating institutions wish to thank 
SLAC for its support and kind hospitality. 
This work is supported by
DOE
and NSF (USA),
NSERC (Canada),
IHEP (China),
CEA and
CNRS-IN2P3
(France),
BMBF and DFG
(Germany),
INFN (Italy),
FOM (The Netherlands),
NFR (Norway),
MIST (Russia), and
PPARC (United Kingdom). 
Individuals have received support from CONACyT (Mexico), 
Marie Curie EIF (European Union),
the A.~P.~Sloan Foundation, 
the Research Corporation,
and the Alexander von Humboldt Foundation.


\begin{thebibliography}{1}

\bibitem{Chau:1993ec}
L.-L. Chau and H.-Y. Cheng, Phys. Lett. {\bf B333},  514  (1994).

\bibitem{Baltrusaitis:1985we}
R.~M. Baltrusaitis {\it et~al.}, Phys. Rev. Lett. {\bf 55},  150  (1985).

\bibitem{Arms:2003ra}
K. Arms {\it et~al.}, Phys. Rev. {\bf D69},  071102  (2004) and 
P. Rubin {\it et~al.}, Phys. Rev. Lett. {\bf 96}, 081802 (2006).

\bibitem{Link:2004vk}
J.~M. Link {\it et~al.}, Phys. Lett. {\bf B618},  23  (2004).

\bibitem{detector}
B. Aubert {\it et~al.}, Nucl. Instrum. Meth. {\bf A479},  1  (2002).

\bibitem{geant4}
S. Agostinelli {\it et~al.}, Nucl. Instrum. Meth. {\bf A506},  250  (2003).

\bibitem{Sjostrand:2000wi}
T. Sj\"ostrand {\it et~al.}, Comput. Phys. Commun. {\bf 135},  238  (2001).

\bibitem{He:2005bs}
Q. He {\it et~al.}, Phys. Rev. Lett. {\bf 95}, 121801 (2005).

\bibitem{Eidelman:2004wy}
S. Eidelman {\it et~al.}, Phys. Lett. {\bf B592},  1  (2004).

\bibitem{Bigi:1994aw}
This calculation assumes ${\cal B}(D^+ \rightarrow K^0_L \pi^+) = 
{\cal B}(D^+ \rightarrow K^0_S \pi^+) $.
Bigi and Yamamoto  pointed out that due to
interference effects these branching fractions could differ by up to 10\%.
I. Y. Bigi and H. Yamamoto, Phys. Lett. {\bf B349}, 363 (1995).

\bibitem{Savage:1991wu}
M.~J. Savage, Phys. Lett. {\bf B257},  414  (1991).

\bibitem{Chiang:2001av}
C.-W. Chiang and J. L. Rosner, Phys. Rev. {\bf D65}, 054007 (2002).

\end{thebibliography}

\end{document}